\documentstyle [12pt] {article}
\makeatletter

\@addtoreset{equation}{section}

\ifcase \@ptsize
\or 
\or 
\fi
\advance\textheight by \topskip

\ifcase \@ptsize
\or 
\or 
\fi

\def\@citex[#1]#2{%
\if@filesw \immediate \write \@auxout {\string \citation {#2}}\fi
\@tempcntb\m@ne \let\@h@ld\relax \def\@citea{}%
\@cite{%
  \@for \@citeb:=#2\do {%
    \@ifundefined {b@\@citeb}%
      {\@h@ld\@citea\@tempcntb\m@ne{\bf ?}%
      \@warning {Citation `\@citeb ' on page \thepage \space undefined}}%
      {\@tempcnta\@tempcntb \advance\@tempcnta\@ne%
      \@tempcntb\number\csname b@\@citeb \endcsname \relax%
      \ifnum\@tempcnta=\@tempcntb 
        \ifx\@h@ld\relax%
          \edef \@h@ld{\@citea\csname b@\@citeb\endcsname}%
        \else%
          \edef\@h@ld{\ifmmode{-}\else--\fi\csname b@\@citeb\endcsname}%
        \fi%
      \else
        \@h@ld\@citea\csname b@\@citeb \endcsname%
        \let\@h@ld\relax%
      \fi}%
    \def\@citea{,\penalty\@highpenalty\,}%
  }\@h@ld%
}{#1}}

\makeatother
\begin{document}
\hfuzz=100pt
\textheight 24.0cm
\topmargin -0.5in
%
%
%
%
\newcommand{\be}{\begin{equation}}
\newcommand{\ee}{\end{equation}}
\newcommand{\bea}{\begin{eqnarray}}
\newcommand{\eea}{\end{eqnarray}}
\begin{titlepage}
\makeatletter
\def \thefootnote {\fnsymbol {footnote}} \def \@makefnmark {
\hbox to 0pt{$^{\@thefnmark }$\hss }}
\makeatother
\begin{flushright}
BONN-HE-93-42\\
October, 1993\\
hep-th/9310169
\end{flushright}
\vspace{1.5cm}
\begin{center}
{ \large \bf Gauss Decomposition, Wakimoto Realisation\\
                and Gauged WZNW Models}\\
\vspace{1cm}
{\large\bf H. Arfaei} \footnote{e-mail: arfaei@IREARN.BITNET }\\
\vspace{.5cm}
Department of Physics\\
Sharif University of Technology\\
P. O. Box 11365-9161\\
Tehran, Iran\\
\vspace{.5cm}
and\\
\vspace{.5cm}
{\large\bf Noureddine Mohammedi}
\footnote{e-mail: nouri@avzw02.physik.uni-bonn.de}
\footnote
{Work supported by the Alexander von Humboldt-Stiftung.}
\\
\vspace{.5cm}
Physikalisches Institut\\
der Universit\"at Bonn\\
Nussallee 12\\ D-53115 Bonn, Germany\\

\baselineskip 18pt
\vspace{1cm}
{\large\bf Abstract}
\end{center}
The implications of gauging the Wess-Zumino-Novikov-Witten (WZNW)
model using the Gauss decomposition of the group elements are explored.
We show that, contrary to standard gauging of WZNW models, this gauging
is carried out by minimally coupling the gauge fields. We find that
this gauging, in the case of gauging an abelian vector subgroup,
differs from the standard one by terms proportional to the field
strength of the gauge fields. We prove that  gauging an abelian
vector subgroup does not have a nonlinear sigma model interpretation.
This is because the target-space metric resulting from the
integration over the gauge fields is degenerate. We demonstrate, however,
that this kind of gauging has a natural interpretation in terms
of Wakimoto variables.  \\
\setcounter {footnote}{0}
\end{titlepage}
\baselineskip 20pt
\section{Introduction}

Wess-Zumino-Novikov-Witten (WZNW) theories provide a unifying
frame-work for a large class of conformally invariant models in two
dimensions. They furnish a Lagrangian realisation for the Kac-Moody
algebra which is the building ground for many other theories [1].
The WZNW
models are each associated with a Lie algebra. An important tool
in establishing the connection between the WZNW and other theories
resides in what is known as the Gauss decomposition.
\par
The essence of this decomposition
consists in experessing locally a generic
element in the Lie group as a product of three matrices: A lower
triangular matrix corresponding to the step operators of the
negative roots, a diagonal matrix corresponding to the Cartan
subalgebra and an upper triangular matrix corresponding to the
step operators of the positive roots.
\par
The gauss decomposition has also been used to find a free
field representation for the WZNW models with arbitrary Kac-Moody
algebra [2].
The generators of the Kac-Moody current algebra are then
realised as functions of a set of free scalar fileds together
with a $\beta\gamma$ system, where $\beta$ is a one-form while
$\gamma$ is a zero-form [2,3].
This realisation is known as the the
Wakimoto representation [3]
and has the consequence that the correlation
functions of the WZNW model are expressed as some linear
combinations of the correlation functions corresponding to these
free fields.
\par
The Gauss decomposition is also at the heart of connectiong
Toda theories (including Liouville theory) to gauged WZNW
models [4-6].
In this decompositon, Toda theories arise as gauge-fixed versions
of gauged WZNW actions upon imposing some constraints on  parts of
the Kac-Moody currents [5].
On the other hand, gauged WZNW models provide
a way of describing the motion of strings on geometrically
nontrivial  backgrounds [7].
These backgrounds are sometimes singular.
It is therefore crucial to understand the gauged WZWN model in the Gauss
decomposition. This is the aim of this paper.
\par
In section two, we find all the global symmetries that respect the
Gauss decompositon and leave the WZNW action invariant. We find that
gauging these symmetries is carried out by minimally coupling the
gauge fields. In other words, all ordinary derivatives are replaced
by covariant derivatives. This is in contrast to standard gauging of the WZNW
model where minimal coupling does not lead to a gauge invariant
action. We show that, when gauging abelian subgroups, the axial gauging
in the Gauss decomposition is identical to the standard gauged WZNW
model. On the other hand, the abelian vector gauging in the Gauss
decomposition differs from the standard gauged WZNW model by a term
proportional to the field strength of the gauge fields.
\par
We show also that the vector gauging in the Gauss decomposition
does not lead to a nonlinear sigma model. This is because the metric
is degenerate. We explicitly prove this in the case of the group
$SL(2,R)$ and $SL(3,R)$. The general case of the group $SL(N,R)$
is finally studied in details.
\par
Although this vector gauging does not have a nonlinear
sigma model model interpretation we show, in section four, that it
does
have a natural interpretation in terms of Wakimoto variables.
We prove that the gauging of $r$ abelian $U(1)$ symmetries
in the Wakimoto representation reduces the number of degrees
of freedom by $2r$. In section
three we give a brief review of the Wakimoto variables
and trace back their origin to the  WZNW model.

\section{Gauging in the Gauss Decomposition}
In this paper, we specify a Lie algebra $\cal {G}$ by its
Cartan subalgebra $\cal{H}$ in a basis $\left\{H_i\,\,,\,\, i=1,\dots ,
r^{{\cal{G}}}\right\}$, where $r^{{\cal{G}}}$ is the rank of $\cal {G}$,
together with a set of step operators $\left\{E_\alpha\,\,,\,\,
E_{-\alpha}\right\}$ corresponding to the set of positive
roots $\alpha\in\Phi^+$. We denote by $M_{\cal G}$ the Lie
group whose Lie algebra is $\cal G$.
The action for the WZNW model defined on the group manifold $M_{\cal G}$
is given by
\bea
I(g)&=&{k\over 8\pi}\int_{\partial B}d^2x\sqrt{-\gamma}\gamma^{\mu\nu}
Tr
\left(g^{-1}\partial_\mu g\right)\left(g^{-1}\partial_\nu g\right)
\nonumber\\
&+&{ik\over 12\pi}\int_{B}d^3y\epsilon^{\mu\nu\rho}Tr
\left(g^{-1}\partial_\mu g\right)\left(g^{-1}\partial_\nu g\right)
\left(g^{-1}\partial_\rho g\right)\,\,\,,
\eea
where $g \in M_{\cal G}$ and $B$ is a three-dimensional manifold
whose boundary is the the two-dimensional surface $\partial B$.
\par
Using the Polyakov-Wiegmann formula [8]
\be
I(g_1g_2)= I(g_1)+I(g_2) + {k\over 4\pi}\int d^2x P^{\mu\nu}_{-}Tr
\left(g^{-1}_1\partial_\mu g_1\right)\left(\partial_\nu g_2 g^{-1}_2\right)
\,\,\,,
\ee
where it is convenient to define the quantities
\be
P^{\mu\nu}_+ = \sqrt{-\gamma}\gamma^{\mu\nu} + i\epsilon^{\mu\nu}
\,\,,\,\,
P^{\mu\nu}_- = \sqrt{-\gamma}\gamma^{\mu\nu} - i\epsilon^{\mu\nu}
\ee
one can show that the WZNW action is invariant under
\be
g(z,\bar z) \rightarrow \bar\Omega(\bar z)g(z,\bar z)\Omega(z)\,\,\,.
\ee
Here $z$ ans $\bar z$ are the complex coordinates on the world-sheet
and our conventions are such that $\gamma^{z\bar z}=1$ and the
antisymmetric tensor is given by $\epsilon^{z\bar z}=i$. The above
transformations generate two commuting copies of a Kac-Moody
algebra at level $k$.
\par
Let us assume that the group element  $g$ can be written according to
the Gauss decomposition
\be
g=g_-g_0g_+ \,\,\,,
\ee
where
\be
g_-=\exp\left(\sum_{\alpha\in \Phi^{+}}\phi^\alpha_- E_{-\alpha}
\right)\,\,,\,\,
g_+=\exp\left(\sum_{\alpha\in \Phi^{+}}\phi^\alpha_+ E_{\alpha}
\right)\,\,,\,\,
g_0=\exp\left(\sum_{i=1}^{r^{\cal G}} \lambda^iH_i
\right)\,\,\,.
\ee
The fields $\phi^\alpha_-\,,\,\phi^\alpha_+$ and $\lambda^i$ are
to be interpreted as the coordinates of the target-space non-linear sigma
model associated to the WZNW model. It is also convenient to think of
$g_-$, $g_+$ and $g_0$ as matrices. The matrices $g_-$ and $g_+$ are,
respectively, lower triangular and upper traingular matrices with units
along the diagonals, while $g_0$ is a diagonal matrix.
\par
A multiple use of the Polyakov-Wiegmann fromula yields
\bea
I(g)&=&I(g_0) + {k\over 4\pi}\int d^2x P_-^{\mu\nu}Tr
\left(g_-^{-1}\partial_\mu g_- g_0
\partial_\nu g_+g^{-1}_+g_0^{-1}\right)\nonumber\\
I(g_0)&=&{k\over 8\pi}\int d^2x \sqrt{-\gamma}\gamma_-^{\mu\nu}Tr
\left(g_0^{-1}\partial_\mu g_0\right)
\left(g_0^{-1}\partial_\nu g_0\right)
\,\,\,.
\eea
In deriving this expression we have made use of the the fact that
\bea
&I(g_-)=I(g_+)=0&\nonumber\\
&Tr\left(g_-^{-1}\partial_\mu g_-\right)\left(\partial_\nu g_0
g^{-1}_0\right)=
Tr\left(g_0^{-1}\partial_\mu g_0\right)\left(\partial_\nu g_+
g^{-1}_+\right)=0\,\,\,.&
\eea
This is because $g_-^{-1}\partial_\mu g_-$ is a lower triangular matrix
with zeros along the diagonal and when multiplied by a lower
triangular matrix or a diagonal matrix gives a lower triangular
matrix with zeros along the diagonal. Therefore the resulting matrix of this
multiplication is always traceless. A similar arguement applies to
$\partial_\mu g_+g^{-1}_+$.
\par
We would like now to look for transformations that are symmetries
of the action (2.7) and which, at the same time, preserve the Gauss
decomposition. Namely, transformations that take $g_-$ and $g_+$
to matrices that are still, respectively, lower and upper triangular with
ones along the diagonals and changes $g_0$ to another
diagonal matrix. There are, indeed, two sets of such transformations.
The first set consists of the transformations
\bea
g_-(z,\bar z)&\rightarrow& U(\bar z)g_-(z,\bar z)U^{-1}(\bar z)
\nonumber\\
g_0(z,\bar z)&\rightarrow& U(\bar z)g_0(z,\bar z)V^{-1}(z)
\nonumber\\
g_+(z,\bar z)&\rightarrow& V( z)g_+(z,\bar z)V^{-1}( z)
\eea
where $U(\bar z)$ and $V( z)$ are two diagonal matrices having
the form
\be
U(\bar z)=\exp\left(\sum_{i=1}^{r_1}u^i(\bar z)H_i\right)
\,\,,\,\,
V(z)=\exp\left(\sum_{i=1}^{r_2}v^i( z)H_i\right)
\ee
and $r_1$ and $r_2$ are smaller or equal to the rank
of the Lie algebra $r^{\cal G}$.
\par
The second set of of transformations is given by
\bea
g_-(z,\bar z)&\rightarrow& \Omega_-(\bar z)g_-(z,\bar z)
\nonumber\\
g_0(z,\bar z)&\rightarrow& g_0(z,\bar z)
\nonumber\\
g_+(z,\bar z)&\rightarrow& g_+(z,\bar z)\Omega_+( z)\,\,\,.
\eea
The matrices $\Omega_-(\bar z)$ and $\Omega_+(z)$ are, respectively, of
the same nature as $g_-$ and $g_+$ and they are written as
\be
\Omega_-(\bar z)=\exp\left(\sum_{\alpha\in \Phi^{+}_1}
\omega_-^{\alpha}(\bar z) E_{-\alpha}\right)
\,\,,\,\,
\Omega_+( z)=\exp\left(\sum_{\alpha\in \Phi^{+}_2}
\omega_+^{\alpha}( z) E_{\alpha}\right)\,\,\,.
\ee
Here $\Phi^{+}_1$ and $\Phi^{+}_2$ are two subsets of
$\Phi^+$, the set of positive roots.
\par
Our aim is to make the matrices $U(\bar z)$, $V(z)$, $\Omega_-(\bar z)$
and $\Omega_+(z)$ depend on both $z$ and $\bar z$. This necessitates the
introduction of  four gauge fields corresponding to the four
different transformations. These we denote by $A_\mu$, $B_\mu$,
$M_\mu$, $N_\mu$ $(\mu=z,\bar z)$ and they correspond, respectively, to the
transformations generated by $U$, $V$, $\Omega_-$, $\Omega_+$. They
are written as
\bea
&A_\mu = \sum_{i=1}^{r_{1}}A_\mu^iH_i\,\,,\,\,
B_\mu = \sum_{i=1}^{r_{2}}B_\mu^iH_i &\nonumber\\
&M_\mu = \sum_{\alpha\in\Phi^{+}_1}M_\mu^{\alpha}
E_{-\alpha}\,\,,\,\,
N_\mu = \sum_{\alpha\in\Phi^{+}_2}N_\mu^{\alpha}
E_{\alpha}&
\eea
and they transform as
\bea
&A_\mu \rightarrow A_\mu - U^{-1}\partial_\mu U\,\,,\,\,
B_\mu \rightarrow B_\mu - V^{-1}\partial_\mu V& \nonumber\\
&M_\mu\rightarrow \Omega_-M_\mu\Omega^{-1}_--
\Omega^{-1}_-\partial_\mu\Omega_-\,\,,\,\,
N_\mu\rightarrow \Omega_+^{-1}N_\mu\Omega_+-
\Omega^{-1}_+\partial_\mu\Omega_+&\,\,\,
\eea
where $U$, $V$, $\Omega_-$ and  $\Omega_+$ are now functions of both
$z$ ans $\bar z$.
The gauged WZNW action in the Gauss decomposition, for both sets of
transformations, is found by minimally coupling the gauge fields (replacing
ordinary derivatives by gauge covariant derivatives). The gauged WZNW
action corresponding to the first  set of transformations in (2.9) is
found
to be
\bea
I(g,A,B)&=&{k\over 8\pi}\int d^2x \sqrt{-\gamma}
\gamma^{\mu\nu}Tr \left(g_0^{-1}D_\mu g_0\right)
\left(g_0^{-1}D_\nu g_0\right)\nonumber\\
&+&{k\over 4\pi}\int d^2x
P_{-}^{\mu\nu}Tr \left(g_-^{-1}D_\mu g_-g_0D_\nu g_+g_+^{-1}g_0^{-1}
\right)\,\,\,,
\eea
where
\bea
D_\mu g_0&=&\partial_\mu g_0 + A_\mu g_0 -g_0 B_\mu\nonumber\\
D_\mu g_-&=&\partial_\mu g_- + A_\mu g_- -g_- A_\mu\nonumber\\
D_\mu g_+&=&\partial_\mu g_+ + B_\mu g_+ -g_+ B_\mu \,\,\,\,.
\eea
On the other hand gauging the second set of transformations in (2.11),
yields
\bea
I(g,M,N)&=&{k\over 8\pi}\int d^2x \sqrt{-\gamma}
\gamma^{\mu\nu}Tr \left(g_0^{-1}\partial_\mu g_0\right)
\left(g_0^{-1}\partial_\nu g_0\right)\nonumber\\
&+&{k\over 4\pi}\int d^2x
P_{-}^{\mu\nu}Tr \left(g_-^{-1}D_\mu g_-g_0D_\nu g_+g_+^{-1}g_0^{-1}
\right)\,\,\,,
\eea
where the covariant derivatives are now given by
\bea
D_\mu g_-&=&\partial_\mu g_- + M_\mu g_-\nonumber\\
D_\mu g_+&=&\partial_\mu g_+ + g_+ N_\mu \,\,\,\,.
\eea
\par
Let us now compare this kind of gauging to what is usually written down
for gauged WZNW models. Let us start by exploring the abelian vector gauging.
This gauging reflects the invariance of the gauged WZNW action under the
transformations
\be
g\rightarrow hgh^{-1}\,\,\,,\,\,\,
{\cal {A}}_\mu\rightarrow {\cal {A}}_\mu - h^{-1}\partial_\mu h\,\,\,,
\ee
where $h$ is a group element of the Cartan subalgebra. The usual gauged WZNW
action is then [9]
\bea
I(g,{\cal {A}})=I(g) &+&{k\over 4\pi}\int d^2x Tr\left(P^{\mu\nu}_-
{\cal {A}}_\mu\partial_\nu g g^{-1} -P^{\mu\nu}_+
{\cal {A}}_\mu g^{-1}\partial_\nu g \right.\nonumber\\
&+&\left.\sqrt{-\gamma}\gamma^{\mu\nu}{\cal {A}}_\mu{\cal {A}}_\nu
-P_-^{\mu\nu}{\cal {A}}_\mu g{\cal {A}}_\nu g^{-1}\right)\,\,\,.
\eea
In the Gauss decomposition, the transformations (2.19)
would correspond to
setting $V=U=h$ and $B_\mu=A_\mu={\cal {A}}_\mu$.
\par
Using the Gauss decomposition for $g$ in the action (2.20),
we get
\bea
I(g,{\cal {A}})=I(g_0)
&+&{k\over 4\pi}\int d^2x
P_{-}^{\mu\nu}Tr \left(g_-^{-1}D_\mu g_-g_0D_\nu g_+g_+^{-1}g_0^{-1}
\right)\nonumber\\
&-&i{k\over 2\pi}\int d^2x
\epsilon^{\mu\nu}Tr \left({\cal {A}}_\mu\partial_\nu g_0g_0^{-1}\right)
\,\,\,,
\eea
where the covariant derivatives are as written in (2.16)
 upon setting
$A_\mu=B_\mu={\cal {A}}_\mu$.
\par
Notice that this last action differs from the gauged WZNW action (2.15)
previously written for the Gauss decomposition. This difference is due to
the peresence of the last term in (2.21) and which can be written as
\be
i{k\over 2\pi}\int d^2x
\epsilon^{\mu\nu} Tr\left(H_iH_j\right)\lambda^i F^j_{\mu\nu}\,\,\,,
\ee
where we have used the expression for $g_0$ in (2.6)
and for $\cal {A}_\mu$
the expression (2.13). Here
$F^i_{\mu\nu}=\partial_\mu {\cal {A}}_\nu^i-\partial_\nu {\cal {A}}_\mu^i$
and the indices $i,j$ span the Cartan subalgebra that we are gauging.
The appearance of terms proportional to the field strength
of the gauge field in gauged nonlinear sigma models have already been
advocated in refs.[10,11].
The inclusion of such terms in gauged WZNW models has
many consequences on the geometrical interpretation of these models [12].
\par
The other interesting abelian gauging is the axial gauging. This is
characterised by the transformations
\be
g\rightarrow hgh\,\,\,,\,\,\,
{\cal {A}}_\mu\rightarrow {\cal {A}}_\mu - h^{-1}\partial_\mu h\,\,\,.
\ee
This is obtained in the Gauss decomposition by setting $V^{-1}=U=h$
and $A_\mu=-B_\mu={\cal {A}}_\mu$. The standard gauged WZNW action
corresponding
to this axial gauging is given by [9]
\bea
I(g,{\cal {A}})=I(g) &+&{k\over 4\pi}\int d^2x Tr\left(P^{\mu\nu}_-
{\cal {A}}_\mu\partial_\nu g g^{-1} +P^{\mu\nu}_+
{\cal {A}}_\mu g^{-1}\partial_\nu g \right.\nonumber\\
&+&\left.\sqrt{-\gamma}\gamma^{\mu\nu}{\cal {A}}_\mu{\cal {A}}_\nu
+P_-^{\mu\nu}{\cal {A}}_\mu g{\cal {A}}_\nu g^{-1}\right)\,\,\,.
\eea
In the Gauss decomposition, this action reduces exactly to the one that is
written in (2.15) upon setting $A_\mu=-B_\mu={\cal {A}}_\mu$.
Finally, the kind of gauging corresponding to our second set
of transformations in (2.11) cannot be compared to the standard gauged
WZNW model. This is because one cannot, in general, gauge a
transformation of the form $g\rightarrow \Omega_1g\Omega_2$
if $\Omega_1$ and $\Omega_2$ are non-abelian and different.
\par
Let us now apply the Gauss decomposition to some simple examples. Since
the axial gauging, in the Gauss decomposition, leads to the usual
gauged WZNW action, we will concentrate here on the vector gauging which
does differ from the usual gauged WZNW model.
\\
\phantom{xxxxxxxxxxxxxxxxxxxxxxxxxxxxxxxxxxxxxxxxxx}
\phantom{xxxxxxxxxxxxxxxxxxxxxxxxxxxxxxxxxxxxxxxxxx}
\\
\noindent
{\large {\bf {The $SL(2,R)$ Case}}}\\
\phantom{xxxxxxxxxxxxxxxxxxxxxxxxxxxxxxxxxxxxxxxxxx}
\\
\noindent
A generic group element $g$ is parametrised through the Gauss
decomposition
\be
g=g_-g_0g_+=
\left(
\begin{array}{cc}1&0\\ \chi&1\end{array} \right)
\left(
\begin{array}{cc}e^{\phi}&0\\ 0& e^{-\phi}\end{array}\right)
\left(
\begin{array}{cc}1&\psi\\ 0&1\end{array} \right)\,\,\,.
\ee
The $U(1)$ vector transformation (obtained by setting $U=V$ in (2.9)) is
chosen
to be generated by the $U(1)$ matrix
\be
H=\left(
\begin{array}{cc}1&0\\ 0& -1\end{array}\right)\,\,\,.
\ee
The gauged WZNW (obtained by settting $B_\mu=A_\mu$ in (2.15))
is given by
\be
I_{sl(2)}={k\over 4\pi}\int d^2x\left[
\sqrt{-\gamma}\gamma^{\mu\nu}
\partial_\mu\phi\partial_\nu\phi
+P_-^{\mu\nu}e^{2\phi}\left(\partial_\mu -2A_\mu\right)\chi
\left(\partial_\nu +2A_\nu\right)\psi\right]\,\,\,.
\ee
In terms of the different fields, the infinitesimal gauge transformations are
\be
\delta\chi=-2\varepsilon\chi\,\,,\,\,
\delta\psi=2\varepsilon\psi\,\,,\,\,
\delta\phi=0\,\,,\,\,\delta A_\mu=-\partial_\mu\varepsilon\,\,\,.
\ee

The integration over the gauge fields $A_\mu$
is a simple Gaussian integration and
its contribution inside the partition fuction is given  by
\be
2\pi\det \left[\left({2k\over\pi}
e^{2\phi}\psi\chi\right)\right]^{-1}
\ee
The logarithm of this determinant
can be calculated using Zeta-function regularisation
and the heat kernel results (see refs.[13]).
The finite contribution comes from the coefficients $a_1$
in the expansion of the heat kernel and is given by
\be
2\pi\det \left[\left({2k\over\pi}
e^{2\phi}\psi\chi\right)\right]^{-1}
={1\over 8\pi}\left(2\phi +\log(\psi\chi)\right)R^{(2)} +const. \,\,\,,
\ee
where $R^{(2)}$ is the two-dimensional scalar curvature.
\par
Therefore, the effective action is
\be
I_{sl(2)}={k\over 4\pi}\int d^2x\sqrt{-\gamma}\left[
\gamma^{\mu\nu}
\partial_\mu\phi\partial_\nu\phi
+{1\over 2}\left(2\phi +\log(\psi\chi)\right)R^{(2)}
\right]\,\,\,.
\ee
This action is still invariant under the local transformations
for $\chi$, $\psi$ and $\phi$ given in (2.28).
We could, therefore, choose
a gauge such that $\chi=1$. This choice leads to a non-propagating
Faddeev-Popov ghosts. Therefore the only remaining fields are $\phi$
and $\psi$. However, $\psi$ has no kenitic term. Thus, the target-space metric
of the corresponding non-linear sigma model is degenerate
(has zero as eigenvalue). Consequently,
this kind of gauging has no non-linear sigma model interpertation
and therefore cannot describe the motion of strings in a curved background.
It has, though, a natural interpretation in terms of Wakimoto variable as seen
in the next section.
\\
\phantom{xxxxxxxxxxxxxxxxxxxxxxxxxxxxxxxxxxxxxxxxxx}
\phantom{xxxxxxxxxxxxxxxxxxxxxxxxxxxxxxxxxxxxxxxxxx}
\\
\noindent
{\large {\bf {The $SL(3,R)$ Case}}}\\
\phantom{xxxxxxxxxxxxxxxxxxxxxxxxxxxxxxxxxxxxxxxxxx}
\\
\noindent
In complete analogy with the $SL(2,R)$ case, we specify a generic
$SL(3,R)$ group element by
\be
g=g_-g_0g_+=
\left(
\begin{array}{ccc}1&0&0\\ \chi_1&1&0\\ \chi_2&\chi_3&1\end{array}\right)
\left(
\begin{array}{ccc}e^{\phi_1}&0&0\\ 0& e^{\phi_2}&0 \\ 0&0&e^{-(\phi_1+\phi_2)}
\end{array}\right)
\left(
\begin{array}{ccc}1&\psi_1&\psi_2\\ 0&1&\psi_3\\
0&0&1\end{array}\right)\,\,\,.
\ee
We choose to gauge the vector transformations generated by the two $U(1)$
matrices
\be
H_1=\left(
\begin{array}{ccc}
1&0&0\\
0&0&0\\
0&0&-1\end{array}\right)\,\,,\,\,
H_2=\left(
\begin{array}{ccc}
0&0&0\\
0&1&0\\
0&0&-1\end{array}\right)
\ee
We need, therefore, to introduce two gauge fields $A^1_\mu$ and $A^2_\mu$
according to (2.13). The different fields transformations are
\bea
&\delta\chi_1=-\left(\varepsilon_1-\varepsilon_2\right)\chi_1\,\,,\,\,
\delta\chi_2=-\left(2\varepsilon_1-\varepsilon_2\right)\chi_2\,\,,\,\,
\delta\chi_3=-\left(\varepsilon_1+2\varepsilon_2\right)\chi_3&\nonumber\\
&\delta\psi_1=\left(\varepsilon_1-\varepsilon_2\right)\psi_1\,\,,\,\,
\delta\chi_2=\left(2\varepsilon_1-\varepsilon_2\right)\psi_2\,\,,\,\,
\delta\chi_3=\left(\varepsilon_1+2\varepsilon_2\right)\psi_3&\nonumber\\
&\delta\phi_1=0\,\,,\,\,\delta\phi_2=0\,\,,\,\,
\delta A_\mu^1=-\partial_\mu\varepsilon_1\,\,,\,\,
\delta A_\mu^2=-\partial_\mu\varepsilon_2\,\,\,,&
\eea
where $\varepsilon_1$ and $\varepsilon_2$ are two infinitesimal gauge
parameters. The gauge invariant action is given by
\bea
I_{sl(3)}&=&{k\over 4\pi}\int
d^2x\sqrt{-\gamma}\gamma^{\mu\nu}\left[
\partial_\mu\phi_1\partial_\nu\phi_1+
\partial_\mu\phi_2\partial_\nu\phi_2+
\partial_\mu\phi_1\partial_\nu\phi_2\right]\nonumber\\
&+&{k\over 4\pi}\int
d^2xP_-^{\mu\nu}\left\{
e^{(\phi_1-\phi_2)}\left(\partial_\mu\chi_1 -(A_\mu^1-A_\mu^2)\chi_1\right)
\left(\partial_\nu\psi_1+(A_\nu^1-A_\nu^2)\psi_1\right)\right.\nonumber\\
&+&e^{(\phi_1+2\phi_2)}
\left(\partial_\mu\chi_3 -(A_\mu^1+2A_\mu^2)\chi_3\right)
\left(\partial_\nu\psi_3+(A_\nu^1+2A_\nu^2)\psi_3\right)
\nonumber\\
&+&e^{(2\phi_1+\phi_2)}
\left[\chi_3\psi_3
\left(\partial_\mu\chi_1 -(A_\mu^1-A_\mu^2)\chi_1\right)
\left(\partial_\nu\psi_1+(A_\nu^1-A_\nu^2)\psi_1\right) \right.\nonumber\\
&+&\left(\partial_\mu\chi_2 -(2A_\mu^1+A_\mu^2)\chi_2\right)
\left(\partial_\nu\psi_2+(2A_\nu^1+A_\nu^2)\psi_2\right)
\nonumber\\
&-&
\chi_3\left(\partial_\mu\chi_1 -(A_\mu^1-A_\mu^2)\chi_1\right)
\left(\partial_\nu\psi_2+(2A_\nu^1+A_\nu^2)\psi_2\right)\nonumber\\
&-&\left.\left.
\psi_3\left(\partial_\mu\chi_2 -(2A_\mu^1+A_\mu^2)\chi_2\right)
\left(\partial_\nu\psi_1+(A_\nu^1-A_\nu^2)\psi_1\right)\right]\right\}
\,\,\,.
\eea

The integration over the gauge fields yields the
following determinant
\bea
&(2\pi)^2\left[\det \left(-{k\over 2\pi}F
\right)\right]^{-1}&
\nonumber\\
&F\equiv
9{s\over r}e^{(3\phi_1+\phi_2)}\left[
suv e^{2\phi_2}\left(u-1\right)\left(v-1\right)
+r^2e^{-\phi_2}+ruve^{-\phi_1}\right]\,\,\,.&
\eea
The gauge invariant variables $u$, $v$, $r$ and $s$ are
defined by
\be
u={\chi_1\chi_3\over\chi_2}\,\,,\,\,
v={\psi_1\psi_3\over\psi_2}\,\,,\,\,
r=\chi_1\psi_1\,\,,\,\,s=\chi_2\psi_2\,\,\,.
\ee
This determinant is regularised in the same manner as in the
$SL(2,R)$ case.
In terms of these new variables, the final
effective action is
\bea
I_{sl(3)}&=&{k\over 4\pi}\int
d^2x\left[\sqrt{-\gamma}\gamma^{\mu\nu}\left(
\partial_\mu\phi_1\partial_\nu\phi_1+
\partial_\mu\phi_2\partial_\nu\phi_2+
\partial_\mu\phi_1\partial_\nu\phi_2\right)\right.\nonumber\\
&+&\left.9s^2F^{-1}e^{2(2\phi_1+\phi_2)}
P_-^{\mu\nu}\partial_\mu u\partial_\nu v\right]
+ {k\over 3\pi}\int d^2x\sqrt{-\gamma}
\log( F)
R^{(2)}\,\,\,.
\eea
Again, the fields $r$ and $s$ have no kinetic terms and
the non-linear sigma model interpretation breaks down.
This kind of gauging has also a representation in terms of
Wakimoto variables.
\\
\phantom{xxxxxxxxxxxxxxxxxxxxxxxxxxxxxxxxxxxxxxxxxx}
\phantom{xxxxxxxxxxxxxxxxxxxxxxxxxxxxxxxxxxxxxxxxxx}
\\
\noindent
{\large {\bf {The $SL(N,R)$ Case}}}\\
\phantom{xxxxxxxxxxxxxxxxxxxxxxxxxxxxxxxxxxxxxxxxxx}
\\
To end this series of examples, we would like to consider
the general case of the group $SL(N,R)$. We will present
a simple arguement which shows that the abelian
vector gauging in the Gauss decomposition leads always to
a degenerate metric. Let us denote the entries of
$g_-$, $g_+$ and $g_0$ by
\bea
\left(g_-\right)_{\alpha\beta}
&=&\left\{\begin{array}{ll}
0\,\,\,\,\,&\alpha<\beta\\
1\,\,\,\,\,&\alpha=\beta\\
\chi_{(\alpha-1)(\alpha - 2)+\beta}\,\,\,\,\,&\alpha>\beta
\end{array}
\right.\nonumber\\
\left(g_+\right)_{\alpha\beta}
&=&\left\{\begin{array}{ll}
\psi_{(\alpha-1)(N-{\alpha\over 2})+\beta-\alpha}\,\,\,\,\,&\alpha<\beta\\
1\,\,\,\,\,&\alpha=\beta\\
0\,\,\,\,\,&\alpha>\beta
\end{array}\right.\nonumber\\
\left(g_0\right)_{ij}
&=&\left\{\begin{array}{ll}
e^{\phi_i}\,\,\,\,\,&i=j\,\,\,,\,\,\,\sum_{l=1}^{N}\phi_l=0\\
0\,\,\,\,\,&i\neq j\end{array}\right.
\eea
The WZNW action has then the general form
\be
I_{sl(n)}=
{k\over 4\pi}\int d^2x\left[\sqrt{-\gamma}
\gamma^{\mu\nu}M^{ij}\partial_\mu\phi_i\partial_\nu\phi_j
+P_-^{\mu\nu}M^{\alpha\beta}
\partial_\mu\chi_\alpha\partial_\nu\psi_\beta\right]
\,\,\,.
\ee
Here $M^{ij}$ $(i,j=1,\dots,r^{\cal G})$ is a constant matrix
while $M^{\alpha\beta}$ $(\alpha,\beta\in\Phi^+)$ is a function
of the fields $\phi_i$, $\chi_\alpha$ and $\psi_\alpha$.
\par
Suppose that we are gauging $r$, $r\leq r^{\cal{G}}$, abelian $U(1)$
symmetries then we need to introduce $r$ gauge fields
$A_\mu^{\tilde i}$, $\tilde i=1,\dots,r$. The gauged WZNW action
in the Gauss decomposition takes then the general form
\bea
I^g_{sl(n)}&=&
{k\over 4\pi}\int d^2x\left[\sqrt{-\gamma}
\gamma^{\mu\nu}M^{ij}\partial_\mu\phi_i\partial_\nu\phi_j\right.
\nonumber\\
&+&\left. P_-^{\mu\nu}M^{\alpha\beta}
\left(\partial_\mu\chi_\alpha - V^\rho_{\tilde i\alpha}A_\mu^{\tilde i}
\chi_\rho\right)
\left(
\partial_\nu\psi_\beta+V^\sigma_{\tilde j\beta}A_\nu^{\tilde j}
\psi_\sigma\right)\right]\,\,\,.
\eea
The quantity $V^\rho_{\tilde i\alpha}$ is a constant matrix and is diagonal in
the indices $\rho$, $\alpha$.
The gauge transformations are
\be
\delta \chi_\alpha=-\varepsilon^{\tilde i}V_{\tilde i \alpha}^\beta\chi_\beta
\,\,\,,\,\,\,
\delta \psi_\alpha=\varepsilon^{\tilde i}V_{\tilde i \alpha}^\beta\psi_\beta
\,\,\,,\,\,\,\delta\phi_i=0\,\,\,,\,\,\,
\delta A_\mu^{\tilde i}=-\partial_\mu\varepsilon^{\tilde i}
\,\,\,.
\ee
and the matrix $M^{\alpha\beta}$ satisfies
\be
-M^{\rho\beta}V_{\tilde i \rho}^\alpha+
M^{\alpha\rho}V_{\tilde i \rho}^\beta-
{\partial M^{\alpha\beta}\over \partial \chi_\rho}
V_{\tilde i \rho}^\sigma\chi_\sigma +
{\partial M^{\alpha\beta}\over \partial \psi_\rho}
V_{\tilde i \rho}^\sigma\psi_\sigma=0
\,\,\,.
\ee
The integration over the gauge fields leads to the nonlinear sigma model
\be
I^g_{sl(n)}=
{k\over 4\pi}\int d^2x\left[\sqrt{-\gamma}
\gamma^{\mu\nu}M^{ij}\partial_\mu\phi_i\partial_\nu\phi_j
+P_-^{\mu\nu}\widetilde M^{\alpha\beta}
\partial_\mu\chi_\alpha\partial_\nu\psi_\beta\right]
\,\,\,,
\ee
where $\widetilde M^{\alpha\beta}$ is given by
\bea
\widetilde M^{\alpha\beta}&=&
M^{\alpha\beta}-(L^{-1})^{\tilde i\tilde j}
M^{\alpha\rho}M^{\tau\beta}V_{\tilde j \rho}^\sigma
V_{\tilde i \tau}^\gamma\chi_\gamma\psi_\sigma\nonumber\\
L_{\tilde i\tilde j}&\equiv& M^{\alpha\beta}
V_{\tilde i \alpha}^\rho
V_{\tilde j \beta}^\sigma\chi_\rho\psi_\sigma\,\,\,.
\eea
Therefore, the integration over the gauge fields yields a
nonlinear sigma model having a target-space metric which is
block diagonal and takes  the form
\be
G^{ab}=\left(\begin{array}{cc}
M^{ij}&0\\
0&\widetilde M^{\alpha\beta}
\end{array}\right)\,\,\,,
\ee
where $a,b=1,\dots,N^2-1$.
Notice, however, that
\be
\widetilde M^{\alpha\beta}V_{\tilde i \alpha}^\sigma\chi_\sigma=
\widetilde M^{\alpha\beta}V_{\tilde i \beta}^\sigma\psi_\sigma=0
\ee
Hence the metric $G^{ab}$ has $2r$ zero eigenvalues and is not invertible.
This is expected since we still have not fixed our gauge. The effect of gauge
fixing will, however, remove (or project out) $r$ null eigenvalues. Therefore
by exhausting all the gauge freedom we are still left with $r$ vanishing
eigenvalues for the metric $G^{ab}$. Hence the metric is still degenerate even
after gauge fixing.
In the next section we will show that the gauging of the abelian vector
subgroup in the Gauss decomposition has a better interpretation in terms of
Wakimoto variables.

\section{The Wakimoto Realisation and the WZNW Model}

We will summarise here the so-called Wakimoto representation of the
WZNW model [3,14].
This is the representation of the Kac-Moody current algebra
in terms of a set of free scalar fields $\phi_i$, $i=1,...,r^{\cal G}$, and
a set of bosonic $\beta\gamma$ system, where $\beta^\alpha$ and
$\gamma_\alpha$ are respectively of spin one and zero. The operator
product expansions of these fields are
\be
\phi_i(z)\phi_j(w)=\kappa^{-1}\delta_{ij}\log( z-w)\,\,\,,\,\,\,
\beta^\alpha(z)\gamma_\sigma(w)=\delta^\alpha_\sigma{1\over z-w}\,\,\,.
\ee
The currents of the Kac-Moody current algebra are then realised as functions
of these free fields.
\par
The Sugawara construction leads to the energy-momentum tensor
\bea
T(z)_{sug}&=&T_{\beta\gamma}(z)+T_{\phi}(z)\nonumber\\
T_{\beta\gamma}(z)&\equiv&:\partial\gamma_\alpha(z)\beta^\alpha(z):\,\,\,,\,\,
\,
\alpha\in\Phi^+\nonumber\\
T_{\phi}(z)&\equiv&{1\over 2}\kappa\delta^{ij}
:\partial\phi_i(z)\partial\phi_j(z):-\rho_i\partial^2\phi^i\,\,\,,
\eea
where $\bf\rho$=${1\over 2}\sum_{\alpha\in\Phi^+}$$\bf\alpha$ is half the
sum of the positive roots and $\kappa=h^{\cal{G}}+k$, where $h^{\cal{G}}$ is
the dual
Coxeter number of $\cal G$.
The central charges coming from the different energy momentum tensors
are given by
\bea
c_{sug}&=&c_{\beta\gamma}+c_\phi\nonumber\\
{k dim{\cal {G}}\over {k+h^{\cal{G}}}}&=&\left(dim{\cal {G}}-r^{{\cal
{G}}}\right)+
\left(r^{\cal G}-{12\rho^2\over {k+h^{\cal{G}}}}\right)\,\,\,,
\eea
where the "strange formula`` of Freudenthal-De Vries,
$12\rho^2=h^{\cal {G}}dim\cal G$, has
been used.
\par
The Sugawara energy-momentum tensor is the Noether current of the action
(in the conformal gauge)
\be
S_{waki}={k\over 2\pi}
\int d^2z\left[\partial_z\gamma_\alpha\beta^\alpha+\kappa \delta^{ij}
\partial_z\phi_i\partial_{\bar z}\phi_j+\rho_i\phi^iR^{(2)}\right]
\,\,\,.
\ee

Since the WZNW action is a realisation of the Kac-Moody algebra, this action
should derive from the WZWN model. This is indeed the case as explained
in great details in ref.[2].
Let us sketch this for the $SL(N,R)$ group. In the Gauss decomposition the
WZWN action takes the general form in (2.40). The spin one fileds,
$\beta^\alpha$,
are obtained from the WZNW action through the change of variables
\be
\beta^\alpha=M^{\alpha\beta}\partial_{\bar z}\psi_\beta
\ee
while the spin zero fields $\gamma_\alpha$ are identified with $\chi_\alpha$.
It is clear then that the last term in the WZNW action (2.40) reproduces
the first term of the Wakimoto action in (3.4). On the other hand, the terms
involving $\phi_i$ in the Wakimoto action are obtained as a combination
of the contribution of $I(g_0)$ (the first term in in the WZNW action (2.40))
and
the contribution arising from the Jacobian of the change of variables made
in (3.5). We refer the reader to ref.[2] for a complete account.

\section{Gauging the Wakimoto Realisation}

The Wakimoto action is invariant under the global transformations
\be
\delta\gamma_\alpha=-\varepsilon^{\tilde i}Q^\beta_{\tilde
i\alpha}\gamma_\beta
\,\,,\,\,
\delta\beta^\alpha=\varepsilon^{\tilde i}Q^\alpha_{\tilde i\beta}\beta^\beta
\,\,,\,\,\delta\phi_i=0\,\,\,,
\ee
where $\tilde i=1,...,r$ and the constant matrix $Q^\beta_{\tilde i\alpha}$
is diagonal in the indices $\alpha,\beta$. Gauging these transformations
requires the introduction of $r$ gauge fields $A_z^{\tilde i}$ transforming
as $\delta A_z^{\tilde i}=-\partial_z\varepsilon^{\tilde i}$. The gauged
action
aquires the form
\be
S^g_{waki}={k\over 2\pi}
\int d^2z\left[\left(\partial_z\gamma_\alpha
-A_z^{\tilde i}Q^\beta_{\tilde i\alpha}\gamma_\beta\right)
\beta^\alpha
+\kappa \delta^{ij}
\partial_z\phi_i\partial_{\bar z}\phi_j+\rho_i\phi^iR^{(2)}\right]
\,\,\,.
\ee
Integrating out the gauge fields is equivalent to imposing the constraints
\be
Q^\beta_{\tilde i\alpha}\gamma_\beta\beta^\alpha=0\,\,\,.
\ee
These constraints would allow us to eliminate $r$ fields of the type $
\beta^\alpha$. Let us, without loss of generality, eliminate the first
$r$ fields of the type $\beta^\alpha$. We have then
\be
\beta_\mu^{\tilde \alpha}=- \gamma^{-1}_{\tilde\beta}
\left(Q^{-1}\right)^{\tilde\beta\tilde\alpha\tilde i}
Q^{\bar \beta}_{\tilde i\bar \alpha}\gamma_{\bar \beta}\beta^{\bar
\alpha}
\,\,\,,
\ee
where the indices $\tilde\alpha,\tilde\beta=1,...,r$ and
$\bar\alpha,\bar\beta=r+1,...,{1\over 2}\left(dim{\cal {G}}-r^{{\cal
{G}}}\right)$
and we have used the fact that $Q^\beta_{\tilde i\alpha}$ is diagonal in the
indices $\alpha,\beta$.
\par
Substituting for $\beta^{\tilde\alpha}$ in the action (4.2),  we obtain
\be
S^g_{waki}={k\over 2\pi}
\int d^2z\left[\partial_z\widehat\gamma_{\bar\alpha}
\widehat\beta^{\bar\alpha}+\kappa \delta^{ij}
\partial_z\phi_i\partial_{\bar z}\phi_j+\rho_i\phi^iR^{(2)}\right]
\,\,\,,
\ee
where the new $\widehat\beta\widehat\gamma$ system is defined by
\be
\widehat\gamma_{\bar\alpha}=-\sum_{\tilde\alpha,\tilde
i=1}^r\left(Q^{-1}\right)
^{\tilde\alpha\tilde\alpha\tilde i}Q^{\bar\alpha}_{\tilde i\bar\alpha}
\log\gamma_{\tilde\alpha}+
\log\gamma_{\bar\alpha}\,\,\,\,,\,\,\,\,
\widehat\beta^{\bar\alpha}=\gamma_{\bar\alpha}\beta^{\bar\alpha}\,\,\,.
\ee
In these last equation there is no summation over the index $\bar\alpha$.
\par
it is crucial to notice that both
$\widehat\gamma_{\bar\alpha}$ and $\widehat\beta^{\bar\alpha}$
are gauge invariant variables. Therefore, any gauge fixing that we could have
imposed on the original $\beta\gamma$ system would
always result in $\left[{1\over 2}\left(dim{\cal {G}}-r^{{\cal {G}}}\right)
-r\right]$
pairs of $\widehat\beta\widehat\gamma$. Therefore, the central charge
corresponding to the energy-momentum tensor of the action (4.5) is given by
\be
c_{\hat\beta\hat\gamma}+c_{\phi}=
\left[\left(dim{\cal {G}}-r^{{\cal {G}}}\right)-2r\right]+
\left(r^{{\cal {G}}}-{12\rho^2{\cal {G}}\over {k+h^{\cal{G}}}}\right)
={kdim{\cal {G}}\over {k+h^{\cal{G}}}}-2r
\,\,\,.
\ee
Hence we see that there are $2r$ (instead of $r$) degrees of freedom
removed by gauging $r$ abelian symmetries in the Wakimoto action.
\par
Let us now take into account
that the Wakimoto action descends from the WZNW model. We will here consider
again the case of the $SL(N,R)$ group.
\par
The gauged WZNW action for the $SL(N,R)$ group is written in (2.41).
Let us choose a
gauge such that
\be
A^{\tilde i}_{\bar z}=0\,\,,\,\,\tilde i=1,\dots,r\,\,\,.
\ee
The Faddeev-Popov action for this gauge fixing is written as
\be
S_{gh}={k\over 2\pi}\int d^2z
\sum_{\tilde i=1}^{r}b_{\tilde i}\partial_{\bar z}c_{\tilde i}
\ee
The action (2.41)  reduces then to
\be
I^g_{sl(n)}=
{k\over 2\pi}\int d^2z\left[M^{ij}\partial_z\phi_i\partial_{\bar z}
\phi_j +
M^{\alpha\beta}
\left(\partial_z\chi_\alpha - V^\rho_{\tilde i\alpha}A_z^{\tilde i}
\chi_\rho\right)
\partial_{\bar z}\psi_\beta\right]+S_{gh}\,\,\,.
\ee
If we change now to the variables $\gamma_\alpha$ and $\beta^\alpha$ as
defined in (3.5) and take into account the contribution due to the Jacobian
(see ref.[2]) then we find
\be
I^g_{sl(n)}=S_{waki}^g+S_{gh}\,\,\,.
\ee
Therefore the gauged Wakimoto action is a gauge-fixed version of the
gauged WZNW model in the Gauss decompositon.
\par
There is, however, a difference between gauging the Wakimoto action and
the gauged WZNW model.
This difference  resides in the following remark. In gauging the
Wakimoto action, the one-form fields $\beta^\alpha$ were allocated
the transformation
\be
\beta^\alpha\rightarrow \beta^\alpha+\varepsilon^{\tilde i}
(z,\bar z)Q^\alpha_{\tilde i\beta}\beta^\beta
\ee
However, if $\beta^\alpha$ is defined in terms of the field $\phi_i$,
$\chi_\alpha$ and $\psi_\alpha$ as in (3.5), then this transformation is
certainly a non-local one. This is because $\beta^\alpha$ depends
on $\partial_{\bar z}\psi_\alpha$ and any local transformation on $
\psi_\alpha$ would appear trough its derivative in $\beta^\alpha$,
which is not what is written above.
\par
The most important point of this section is that we are able
to give an interpretation in terms of Wakimoto variables to the
gauged WZNW model in the Gauss decomposition. As  seen before, this
kind of gauging does not have a nonlinear sigma model interpretation.

\section{Conclusions}

In the Wess-Zumino-Novikov-Witten models, there are only two possible
symmetries that can be gauged: The vector gauging $g\rightarrow hgh^{-1}$,
where $h$ can be abelian or non-abelian group element, and the axial gauging
$g\rightarrow hgh$, where in this case $h$ must be abelian.
\par
If, however, $g$ admits a Gauss decomposition, $g=g_-g_0g_+$,
then there are two different
symmetries which respect the Gauss decomposition and which can be gauged.
The first such symmetry is an abelian one $g\rightarrow UgV$ and contains
the vector symmetry $(V=U^{-1})$ and the axial symmetry $(V=U)$. The second
symmetry is a non-abelian one $ g\rightarrow \Omega_-g\Omega_+$,
where $\Omega_-$ and $\Omega_+$ are two group elements corresponding
respectively to the lowering and raising step operators.
\par
Our results show that the gauging of these two kinds of symmetries
in the Gauss decomposition is carried out by minimally coupling
the gauge fields. In the case of the abelian gauging we found that
the axial gauging in the Gauss decomposition is identical to the
standard axial gauging of the WZNW model. The vector gauging in the Gauss
decomposition, on the other hand, differs from the standard vector gauging
of the WZNW model by by a term proportional to the field stength of
the gauge fields. The theory resulting from this
vector gauging is still conformally invariant and has a nice interpretation
in terms of Wakimoto variables. However, the resulting nonlinear sigma model
has a degenerate metric.
\par
We conclude therefore that the gauging of the WZNW models depend very
much on the local parametrisation of the group elements. In other
words, once the nonlinear sigma model corresponding to the WZWN model
is written down, the gauging of the isometries of this sigma model
may not lead to a theory which describes the motion of strings
on nontrivial backgrounds.
\par
There are still interesting issues to be explored in this context.  First
of all, the supersymmetric case is straightforward. It seems to us that
the duality between the vector gauging and the abelian gauging  found
in [15] is broken in the Gauss decomposition. This needs further
investigations. Furthermore, Sfetsos and Tseytlin [16] have also considered a
chiral gauging of the WZNW model (which differs from the standard
gauging) and it would be of interest to explore the connection
between our gauging in the Gauss decomposition and this chiral gauging.
Finally, since the Gauss decomposition is used to connect
gauged WZNW models and Toda theories, we feel that it is necessary
to re-examine the non-abelian gauging in the Gauss decomposition. This
would then shed some light on the origin of the exactly conformally action
for non-abelian Toda theories found in ref.[17]. We hope to return to these
points in the near future.

\vspace{0.5cm}
\paragraph{Acknowledgements:}
We would like to thank Jose
Figueroa-O'Farrill, Ian Jack and Werner Nahm
for  many useful discussions. One of us (H. A.)
would like to thank the Max Planck Institut (MPI)
and the Physikalisches Institut in Bonn for
their hospitality. The research of (N. M.)
is  supported  by the Alexander von
Humboldt-Stiftung.


\begin{thebibliography}{99}
\bibitem{1}
E, Witten, Comm. Math. Phys. 92 (1984) 483.
\bibitem{2}
A. Gerasimov, A. Marshakov, A. Morozov. M. Olshanetsky and
S. Shatashvili, Int. J. Mod. Phys. A5 (1990) 2495.
\bibitem{3}
M. Wakimoto, Comm. Math. Phys. 104 (1986) 605.
\bibitem{4}
L. O'Raifeartaigh and A. Wipf, Phys. Lett. B251 (1990) 361.
\bibitem{5}
J. Balog, L. Feh\'er, P. Forg\'acs, L. O'Raifeartaigh, and A. Wipf,
Ann. Phys. (New York) 203 (1990) 76.
\bibitem{6}
J. Balog, L. Feh\'er, P. Forg\'acs, L. O'Raifeartaigh,  and A. Wipf,
Phys. Lett. B227 (1989) 214.
\bibitem{7}
E. Witten, Phys. Rev. D44 (1991) 314.
\bibitem{8}
A. M. Polyakov and P. B. Wiegmann, Phys. Lett. Lett. B131 (1983) 121.
\bibitem{9}
P. Di Vecchia and P. Rossi, Phys. Lett. B140 (1884) 344;
\newline
P. Di Vecchia, B. Durhuus and J. Petersen, Phys. Lett. B144 (1884);
\newline
D. Karabali, Q.-H. Park, H. J. Schnitzer and Z. Yang, Phys. Lett. B216 (1989)
307; \newline
H. J. Schnitzer, Nucl. Phys. B324 (1989) 412;\newline
D. Karabali and H. J. Schnitzer, Nucl. Phys. B329 (1990) 649;
\newline
K. Gawedzki and A. Kupianen, Phys. Lett. B215 (1988) 119; Nucl. Phys. B320(FS)
(1989) 649.
\bibitem{10}
I. Jack, D. R. T. Jones, N. Mohammedi and H. Osborn, Nucl. Phys. B332 (1990)
359.
\bibitem{11}
C. M. Hull and B. Spence, Phys. Lett. B232 (1989) 204
\bibitem{12}
N. Mohammedi, 'Modified Black Holes in Two Dimensional Gravity',
Phys. Lett. B281 (1992) 36;
\newline
N. Mohammedi, 'A large Class of Spacetimes from the $SL(2,R)/U(1)$
Coset Model', Mod. Phys. Lett. A8 (1993) 1125;
\newline
N. Mohammedi, 'Naked Singularities in Four-dimensional String
Backgrounds', Phys. Rev. D (to appear).
\bibitem{13}
E. Corrigan, P. Goddard, H. Osborn and S. Templeton,
Nucl. Phys. B159 (1979) 469;
\newline
S. W. Hawking, Comm. Math. Phys. 55 (1977) 133;
\newline
I. Jack, H. Osborn, Nucl. Phys. B207 (1982) 474.
\bibitem{14}
H. Awata, A. Tsuchiya and Y. Yamada, Nucl. Phys. B365 (1991) 680.
\bibitem{15}
E. Kiritsis, Mod. Phys. Lett. A6 (1991) 2871.
\bibitem{16}
K. Sfetsos and A. A. Tseytlin, 'Chiral gauged WZNW models and
heterotic string backgrounds' preprint CERN-TH.6962/93 (1993).
\bibitem{17}
I. Jack, D. R. T. Jones and J. Panvel, 'Quantum Non-abelian
Toda Field Theories', Liverpool preprint LHT315 (1993).
\end{thebibliography}
\end{document}